\input harvmac.tex

\input epsf.tex


\def\figin{\epsfcheck\figin}\def\figins{\epsfcheck\figins}
\def\epsfcheck{\ifx\epsfbox\UnDeFiNeD
\message{(NO epsf.tex, FIGURES WILL BE IGNORED)}
\gdef\figin##1{\vskip2in}\gdef\figins##1{\hskip.5in}
\else\message{(FIGURES WILL BE INCLUDED)}%
\gdef\figin##1{##1}\gdef\figins##1{##1}\fi}
\def\DefWarn#1{}
\def\figinsert{\goodbreak\midinsert}
\def\ifig#1#2#3{\DefWarn#1\xdef#1{fig.~\the\figno}
\writedef{#1\leftbracket fig.\noexpand~\the\figno}%
\figinsert\figin{\centerline{#3}}\medskip\centerline{\vbox{\baselineskip12pt
\advance\hsize by -1truein\noindent\footnotefont{\bf
Fig.~\the\figno:} #2}}
\bigskip\endinsert\global\advance\figno by1}



\def\stokesparameter{\gamma}
\def\wshifted{w_s}


\lref\DrummondAUA{
  J.~M.~Drummond, G.~P.~Korchemsky and E.~Sokatchev,
  Nucl.\ Phys.\  B {\bf 795}, 385 (2008)
  [arXiv:0707.0243 [hep-th]].
}

\lref\AldayHR{
  L.~F.~Alday and J.~M.~Maldacena,
  JHEP {\bf 0706}, 064 (2007)
  [arXiv:0705.0303 [hep-th]].
}

\lref\AldayYW{
  L.~F.~Alday and R.~Roiban,
  Phys.\ Rept.\  {\bf 468}, 153 (2008)
  [arXiv:0807.1889 [hep-th]].
}

\lref\DrummondAUA{
  J.~M.~Drummond, G.~P.~Korchemsky and E.~Sokatchev,
  Nucl.\ Phys.\  B {\bf 795}, 385 (2008)
  [arXiv:0707.0243 [hep-th]].
}

\lref\BrandhuberYX{
  A.~Brandhuber, P.~Heslop and G.~Travaglini,
  Nucl.\ Phys.\  B {\bf 794}, 231 (2008)
  [arXiv:0707.1153 [hep-th]].
}

\lref\OoVa{
  H.~Ooguri and C.~Vafa,
  Phys.\ Rev.\ Lett.\  {\bf 77}, 3296 (1996)
  [arXiv:hep-th/9608079].
}

\lref\Brand{
  C.~Anastasiou, A.~Brandhuber, P.~Heslop, V.~V.~Khoze, B.~Spence and G.~Travaglini,
  arXiv:0902.2245 [hep-th].
}

\lref\SeibergNS{
  N.~Seiberg and S.~H.~Shenker,
  Phys.\ Lett.\  B {\bf 388}, 521 (1996)
  [arXiv:hep-th/9608086].
}

\lref\FG{
  V.V. Fock, A.B. Goncharov,
  arXiv:math/0311149}

\lref\DeVegaXC{
  H.~J.~De Vega and N.~G.~Sanchez,
  Phys.\ Rev.\  D {\bf 47}, 3394 (1993).
}

\lref\GaiottoCD{
  D.~Gaiotto, G.~W.~Moore and A.~Neitzke,
  arXiv:0807.4723 [hep-th].
}

\lref\GMNtwo{
  D.~Gaiotto, G.~W.~Moore and A.~Neitzke, to appear
}

\lref\HitchinVP{
  N.~J.~Hitchin,
  Proc.\ Lond.\ Math.\ Soc.\  {\bf 55}, 59 (1987).
}


\lref\JevickiAA{
  A.~Jevicki, K.~Jin, C.~Kalousios and A.~Volovich,
   arXiv:0712.1193 [hep-th].
}

\lref\MaldacenaIM{
  J.~M.~Maldacena,
  Phys.\ Rev.\ Lett.\  {\bf 80}, 4859 (1998)
  [arXiv:hep-th/9803002].
}

\lref\ReyIK{
  S.~J.~Rey and J.~T.~Yee,
  Eur.\ Phys.\ J.\  C {\bf 22}, 379 (2001)
  [arXiv:hep-th/9803001].
}

\lref\DrummondAQ{
  J.~M.~Drummond, J.~Henn, G.~P.~Korchemsky and E.~Sokatchev,
  arXiv:0803.1466 [hep-th].
}

\lref\PohlmeyerNB{
  K.~Pohlmeyer,
  Commun.\ Math.\ Phys.\  {\bf 46}, 207 (1976).
}

\lref\BernAP{
  Z.~Bern, L.~J.~Dixon, D.~A.~Kosower, R.~Roiban, M.~Spradlin, C.~Vergu and A.~Volovich,
  Phys.\ Rev.\  D {\bf 78}, 045007 (2008)
  [arXiv:0803.1465 [hep-th]].
}

\lref\BernIZ{
  Z.~Bern, L.~J.~Dixon and V.~A.~Smirnov,
  Phys.\ Rev.\  D {\bf 72}, 085001 (2005)
  [arXiv:hep-th/0505205].
}

\lref\AM{
  L.~F.~Alday and J.~Maldacena, To appear.
}

\lref\JevickiUZ{
  A.~Jevicki and K.~Jin,
  arXiv:0903.3389 [hep-th].
}

\lref\KruczenskiFB{
  M.~Kruczenski,
  JHEP {\bf 0212}, 024 (2002)
  [arXiv:hep-th/0210115].
}

\lref\SeibergNS{
  N.~Seiberg and S.~H.~Shenker,
  Phys.\ Lett.\  B {\bf 388}, 521 (1996)
  [arXiv:hep-th/9608086].
}

\Title{\vbox{\baselineskip12pt \hbox{} \hbox{
} }} {\vbox{\centerline{Minimal surfaces in $AdS $ and the
  }
\centerline{ eight-gluon scattering amplitude } \centerline{ at
strong coupling } }}
\bigskip
\centerline{Luis F. Alday and Juan Maldacena }
\bigskip
\centerline{ \it  School of Natural Sciences, Institute for
Advanced Study} \centerline{\it Princeton, NJ 08540, USA}

\vskip .3in \noindent
In this note we consider minimal surfaces in three dimensional
anti-de Sitter space that end at the $AdS$ boundary on a polygon
given by a sequence of null segments. The problem can be reduced
to a certain generalized Sinh-Gordon equation and to $SU(2)$
Hitchin equations. The mathematical problem to be  solved arises
also in the context of the moduli space of certain three
dimensional supersymmetric theories.  We can use explicit results
available in the literature in order to find the explicit answer
for the area of a surface that ends on a eight-sided null Wilson
loop. Via the gauge/gravity duality this can also be interpreted
as a certain eight-gluon scattering amplitude at strong coupling
for a special kinematic configuration.


 \Date{ }

\newsec{Introduction }

Recently there has been some interest in
Wilson loops that consist of a sequence of light-like segments.
 These are interesting
for several reasons. First, they are a simple subclass of Wilson loops
 which depend on a finite number
of parameters, the positions of the cusps. Second, they are Lorentzian objects with no obvious
Euclidean counterpart. Finally, it was shown that they are connected to scattering amplitudes
in gauge theories \refs{\AldayHR,\DrummondAUA,\BrandhuberYX,\BernAP,\DrummondAQ},
for a review see \AldayYW .

In this note we study these Wilson loops at strong coupling by using the gauge/string duality. One is then led to compute the area of minimal surfaces in $AdS$ \MaldacenaIM \ReyIK . We consider a special class of
 null Wilson loops which can be embedded in  a two dimensional subspace, which we can take as an
$R^{1,1}$ subspace of the boundary of $AdS$. For these loops, the string worldsheet lives in an
$AdS_3$ subspace of the full $AdS_d$ space, $d\geq 3$.

 In order to analyze the problem one can use a  Pohlmeyer type reduction
 \refs{\PohlmeyerNB,\DeVegaXC,\JevickiAA,\JevickiUZ}.
 This maps the problem
 of strings moving in $AdS_3$ to a problem involving a single field $\alpha$ which obeys
 a generalized Sinh-Gordon equation.

  The same mathematical problem
  appears in the study of $SU(2)$ Hitchin equations \HitchinVP . Interestingly,
   these Hitchin equations also appear in the study of the supersymmetric
 vacua of certain gauge theories \refs{\GaiottoCD,\GMNtwo}. This connection
 is specially useful because the authors of \refs{\GaiottoCD,\GMNtwo}
  have studied this problem,
  exploiting its integrability, and have proposed a method for finding the answer.
  For the simplest case, the metric in moduli space for the corresponding field theory problem is
  known \refs{\OoVa,\SeibergNS,\GaiottoCD,\GMNtwo}.
   These results can be used to compute the area for the simplest non-trivial case which is the
   eight sided Wilson loop.

This note is organized as follows. In section  2 we describe the
interplay between classical strings on $AdS_3$ and the generalized
Sinh-Gordon model. In section 3 we describe several features of
the solutions at hand and in section 4 we give the full answer for
the case of the null Wilson loop with eight sides.

Note: A much more detailed exposition of the material reported
here will be presented in a forthcoming publication \AM .

\newsec{Sinh-Gordon model from strings on $AdS_3$}

Classical strings in $AdS$ spaces can be described by a  reduced
model which depends only on physical degrees of freedom. In terms
of embedding coordinates, $Y^\mu$ in $R^{2,2}$, with $Y^2=-1$,
the conformal gauge equations of motion and   Virasoro constraints
are
\eqn\eom{\partial \bar \partial \vec{Y}-(\partial \vec{Y}.\bar{\partial} \vec{Y})\vec{Y}=0 ~,~~~~~~
  \partial \vec{Y}.\partial \vec{Y}=\bar \partial \vec{Y}.\bar \partial \vec{Y}=0 }
where we parametrize the world-sheet in terms of  complex
variables $z$ and $\bar{z}$. For the case of $AdS_3$ the above
system can be reduced to the generalized sinh-Gordon model
\refs{\PohlmeyerNB,\DeVegaXC,\JevickiAA,\JevickiUZ}. We define
\eqn\sinhg{\eqalign{e^{2 \alpha(z,\bar{z})} &= {1 \over 2 } \partial \vec{Y}.\bar \partial \vec{Y},~~~~~
N_a =  { e^{-2 \alpha} \over 2 } \epsilon_{abcd}Y^b \partial Y^c \bar \partial Y^d ~,
\cr p&=  -{1 \over 2}  \vec{N}.\partial^2 \vec{Y}~,~~~~~~~ \bar p= {1 \over 2}  \vec{N}.{ \bar \partial}^2 \vec{Y}
}}
Then, as a consequence of \eom\ it can be shown that $p=p(z)$ is a
holomorphic function and that $\alpha(z,\bar{z})$ satisfies the
generalized Sinh-Gordon equation
\eqn\gensinh{\partial \bar \partial \alpha(z,\bar{z})-e^{ 2 \alpha(z,\bar{z}) }+|p(z)|^2 e^{-2 \alpha(z,\bar{z})}=0}
The area of the world-sheet is simply given by
\eqn\area{ A =4 \int d^2z  e^{2 \alpha }}
For solutions relevant to this note this area is divergent and we will need to regularize it.

Given a solution of the generalized sinh-Gordon model, one can reconstruct a classical string world-sheet in $AdS_3$ by solving two auxiliary linear problems (which we will denote as left and right)
\eqn\LS{\eqalign{ \partial \psi^L_{\alpha,a}+(B_z^L)_{\alpha}^{~\beta}\psi^L_{\beta,a}=0,~~~~~
\bar{\partial} \psi^L_{\alpha,a}+(B_{\bar z}^L)_{\alpha}^{~\beta}\psi^L_{\beta,a}=0 \cr
\partial \psi^R_{\dot \alpha,\dot a }+(B_z^R)_{\dot \alpha}^{~\dot \beta}\psi^R_{\dot \beta, \dot a } =0,~~~~~\bar{\partial} \psi^R_{\dot \alpha,\dot a }+(B_z^R)_{\dot \alpha}^{~\dot \beta}\psi^R_{\dot \beta, \dot a } =0
}}
where the $SL(2)$ left and right flat connections are given by
\eqn\FGhybrid{\eqalign{B_z^L=\left(\matrix{{1 \over 2} \partial \alpha &- {e^{  \alpha}  }\cr-{e^{-  \alpha} p(z) }  & -{1 \over 2} \partial \alpha } \right),~~~~~ B_{\bar{z}}^L=\left(\matrix{-{1 \over 2} \bar{\partial} \alpha & -{e^{-  \alpha } {\bar p }(\bar{z})  }\cr- e^{ \alpha}    & {1 \over 2} \bar{\partial} \alpha }\right) \cr
B_z^R=\left(\matrix{-{1 \over 2} \partial \alpha &{e^{-  \alpha} p(z)  }   \cr -{e^{  \alpha}  } & {1 \over 2} \partial \alpha }\right),~~~~~
B_{\bar{z}}^R=\left(\matrix{{1 \over 2} \bar{\partial} \alpha & -{e^{  \alpha} }  \cr{e^{-  \alpha} {\bar p }(\bar{z}) }  & -{1 \over 2} \bar{\partial} \alpha }\right) }}
Internal $SL(2)_L \times SL(2)_R$ indices  $\alpha, \dot \alpha$
denote rows and columns of these connections, while the indices
$a$ and $\dot{a}$ denote the two different linearly independent
solutions of each linear problem \LS\ . The space-time
isometry group $SO(2,2)=SL(2) \times SL(2)$ acts on these indices.
 We require that the two  pairs  of solutions obey the normalization
 condition
\eqn\normalization{\psi^L_a \wedge \psi^L_b \equiv
\epsilon^{\beta \alpha } \psi^L_{\alpha,a} \psi^L_{\beta,b}=
\epsilon_{a b},~~~~~\psi^R_{\dot a} \wedge \psi^R_{\dot b}=\epsilon_{\dot a \dot b}}
Once a pair of solutions has been found, the explicit form of  the
space-time coordinates $Y_{a \dot a}(z,\bar{z})$ is given by a
particular bilinear combination of the left and right solutions
\eqn\inversemap{\eqalign{  Y_{a \dot a } =
\left(\matrix{Y_{-1}+Y_{2}& Y_1-Y_0\cr Y_1 + Y_0 & Y_{-1}-Y_{2} }\right)_{a,\dot{a}}=
\psi^L_{\alpha,a} M_1^{\alpha \dot \beta} \psi^R_{\dot \beta ,\dot{a}} ~,~~~~~~~~~M_1^{\alpha \dot
\beta } =  \pmatrix{ 1 & 0 \cr 0 & 1 }
}}
\bigskip
It turns out that the left connection $B^L$ can be promoted to a
family  of flat connections by introducing a spectral parameter
\eqn\condec{B_z=A_z+\Phi_z \rightarrow B_z(\zeta)=A_z+{ 1 \over \zeta }
\Phi_z,~~~~~~B_{\bar z}=A_{\bar z}+\Phi_{\bar z} \rightarrow B_{\bar z}(\zeta)=A_{\bar z}+{ \zeta} \Phi_{\bar z}}
where we have decomposed the connection into its diagonal  part
$A_z$ and off diagonal part $\Phi_z$. Actually, both, left and
right connections can be simply obtained (up to a constant gauge
transformation) from $B(\zeta)$ by setting $\zeta=1$ or $\zeta=i$
respectively. The zero curvature condition for $B(\zeta)$ can be
rephrased as
\eqn\hitchineq{\eqalign{D_{\bar z} \Phi_z&=D_{z} \Phi_{\bar z}=0,~~~~~F_{z \bar{z}}+[\Phi_z,\Phi_{\bar z}]=0\cr
D_\mu \Phi&=\partial_\mu \Phi+[A_\mu,\Phi],~~~~~F_{\mu \nu}=\partial_\mu A_\nu-\partial_\nu A_\mu+[A_\mu,A_\nu]}}
 These are the Hitchin equations, which arise by dimensional
  reduction of the four dimensional self-duality condition
  (instanton equations) to two dimensions. $A$ has the interpretation
   of a gauge connection in two dimensions and $\Phi$ is a Higgs field.
    In our case we have the  Hitchin equations for $SU(2)$.

After we compute the area we can relate these results to results
for Wilson loops or amplitudes in ${\cal N}=4 $ super Yang Mills
as
 \eqn\reswla{ \langle W \rangle \sim {\rm Amplitude} \sim e^{ - { R^2
\over 2 \pi \alpha' } ({\rm Area } ) } ~,~~~~~~~ {R^2 \over \alpha
' } = \sqrt{\lambda} = \sqrt{ g^2 N }
}
  where the area is computed
by setting the $AdS$ radius to one  and we have embedded $AdS_3$
appropriately in $AdS_5$. For other field theories with gravity
duals we should use the corresponding expression for
$R^2/\alpha'$.

\newsec{Finding the surface and computing its area}

Since $p(z)$ is a holomorphic function we can simplify the
generalized sinh-Gordon equation by defining a new variable
$dw=\sqrt{p(z)}dz$. In the $w-$plane the equation takes the form
\eqn\wsinh{
 \partial_w \bar \partial_{\bar w} \hat \alpha  - e^{2 \hat \alpha } + e^{ - 2 \hat \alpha } =0 ~,~~~~~~~~~~
 \hat \alpha \equiv \alpha - { 1 \over 4 } \log p \bar p
 }
 The expression for the area
  \area\
  becomes $A = 4 \int d^2 w e^{2 \hat \alpha } $.
 Note that $w$ has branch cuts where $p$ has zeros. So, we locally simplify the equation but
 we complicate the space on which the equation is defined.

Let us first write the solution for the four sided polygon
obtained in \refs{\KruczenskiFB,\AldayHR}. In this case $p=1$,
$w=z$ and $\alpha =0$.
 The connections \FGhybrid\ are constant. Up to a constant gauge
transformation the two solutions can be chosen as
\eqn\twosol{
  \psi_{+}^L = \pmatrix{ e^{    w + \bar w  } \cr 0 }
 ~,~~~~~~~ \psi^L_{-} = \pmatrix{ 0 \cr e^{-   (w + \bar w) }  }
 }
The $\psi_\pm^R$ are the same but with $w + \bar w \to { w - \bar
w \over i } $. Some of these solutions diverge when $|z| \to
\infty$. This implies that the AdS embedding coordinates
\inversemap\ are diverging, which means that we are going to the
$AdS$ boundary. Different components of the $Y$ coordinates in
\inversemap\ are different combinations of the solutions
$\psi^{L,R}_{\pm}$. The solutions that diverges determine a point
on the $AdS$ boundary. This point depends on whether the solution
that diverges is $\psi_+$ or $\psi_-$. Thus, in each of the four
quadrants of the $w$ plane we approach a different point on the
$AdS$ boundary. Each of these points is a cusp. When we change
quadrants the solutions change dominance only for the left problem
or only for the right problem. This implies that two consecutive
cusps are separated by a null line.

Consider now the case in which $p(z)$ is a generic  polynomial of
degree $n-2$, $p \sim z^{ n-2} + \cdots $. In this case $w \sim
z^{n/2}$ for large $z$. As we go once around the $z$ plane we go
around  $n/2$ times in the $w$ plane. Since we expect that the
solution near each cusp is similar to the solution for the four
sided polygon, we demand that $\hat \alpha \to 0$ as $|z|$ goes to
infinity. As a result, the solutions for large $w$ become
approximately as in  \twosol . The problem displays the Stokes
phenomenon. Namely, an exact solution is given in terms of a
linear combination of each of the two solutions \twosol , with
coefficients that change as we move between Stokes sectors. For
instance, consider the left problem in the
  large $|w|$ region, such as $Im(w)<0$. Within that region \twosol\ is a good
  approximate solution. Let us consider now what happens as we cross the line where $w$ is real and
positive, with very large $|w|$.
  The solution that decreases as $Re(w ) \to + \infty$ is accurately given by
\twosol\ and it is the same on both sides of the line. On the
other hand, the large solution will have a jump in its small
solution component. More precisely, we choose a basis of solutions
which has the asymptotic form in   \twosol \ in one Stokes sector.
We denote these two exact solutions as $\psi_\pm^{\rm before} $.
After we cross the Stokes line, we enter into a new Stokes sector.
We can now choose another basis of solutions which has the
asymptotic form in \twosol\ in this new sector. We denote these
two exact solutions as  $\psi_\pm^{\rm after}$. These two sets of
solutions should be related by a simple linear transformation. In
fact we have
 \eqn\matris{
 \psi_{a}^{\rm before } = S_a^{~b} \,
 \psi_{b}^{ \rm after }  ~,~~~~~~~~~~~~~~~
  S(\stokesparameter) = \pmatrix{ 1 & \stokesparameter \cr 0 & 1 }
 }
 The Stokes matrix acts on the target space  $SL(2)$ index, $a= \pm $,  of the solutions.
 In other words $\psi_\pm ^{\rm before} $
 has a new asymptotic expression in the new sector.
 A pair of  exact solutions has  an approximate expression in each sector which is
 given by a
 linear combination of the two solutions in \twosol . These coefficients also change by
 multiplication of the matrix in \matris\ as we change sectors.

 The right problem will display a similar phenomenon. The discontinuities will be then characterized by Stokes parameters $\gamma_i^L$ and $\gamma_i^R$, where $i=1,..,n$ runs over the Stokes lines. In general, the value of the $\gamma's$ depends on the full exact solution and cannot be computed purely at large $|w|$.

 The Stokes lines for the left linear problem \FGhybrid\
  are at  $Re(w)=0$ and the ones for the right problem are at $Im(w)=0$. In addition, within
  each Stokes sector the solutions in \twosol\ exchange dominance at the anti-Stokes lines,
  which for the left problem are at $Im(w) =0$ and for the right problem are at $Re(w)=0$. In conclusion, in each quadrant of the
  $w$ plane the solution diverges and it approaches a point on the $AdS_3$ boundary, the location
  of a cusp. As we change from one quadrant to the next we encounter an anti-Stokes line of one
  of the two problems, say the left problem,
   and now a different solution is diverging, leading to a different point
  on the $AdS$ boundary. At this anti-Stokes line the right problem continues to have the same
  dominant solution. This implies that the two points corresponding to two consecutive
  quadrants are related by only a left target space $SL(2)$ transformation and are thus lightlike
  separated.
   In conclusion, a polynomial $p$ of degree $n-2$ leads to $ 2 n$ different quadrants in the $w$
   plane and thus $ 2n $ cusps.  We can  form $2(n-3)$ different
   spacetime cross ratios out of these points. This number coincides with the number of
   non-trivial  real parameters in the
   coefficients of a polynomial of degree $n-2$, $p = z^{n-2} + c_{n-4} z^{n-4 } + \cdots +
   c_0 $, where we used scalings and translations to bring the polynomial to this form. The
   first case where we can form cross ratios is $n=4$ which corresponds to the
   octagon. For Wilson loops made of null segments in $R^{1,3}$
   the first non-trivial case is the six sided polygon.

  One can write an elegant formula for the left spacetime cross ratios purely in terms of
  the solutions of the left linear problem in \FGhybrid .
  Choosing Poincare coordinates for $AdS_3$, $ds^2 = (dx^+ dx^- + dr^2 )/r^2 $, we can
  write the spacetime  cross ratios for four points $x^+_i$, not necessarily consecutive, as
\eqn\formcross{ { x^+_{12} \,  x^+_{34} \over x^+_{13} \, x^+_{24}
} = { s^L_1 \wedge s^L_2 \,s^L_3\wedge s^L_4 \over s^L_1 \wedge
s^L_3 \, s^L_2 \wedge s^L_4 }\equiv \chi(\zeta=1) } where $s_i$ is
the small solution at the cusp $i$. The $s_i$  are well defined up
to a rescaling, which cancels in \formcross . We have similar
expressions for the right problem, which can be obtained by
introducing the spectral parameter and setting $\zeta =i$. The
expressions in the right hand side are the cross ratios introduced
in \refs{\GMNtwo,\FG}.

\bigskip

A very similar mathematical problem was considered by  Gaiotto, Moore and Neitzke in a very different context \GaiottoCD  . Their motivation was the study of the Hyperkahler moduli space of certain three dimensional gauge theories with ${\cal N}=4$ susy. Those theories can arise from wrapping $D4-$branes on Riemann surfaces.
 The classical Higgs branch moduli space of vacua of these theories is given by
 the moduli space of the Hitchin equations on the corresponding
Riemann surface. The moduli space is parametrized by the
coefficients of the polynomial $p=\prod_{i=1}^{n-2}(z-z_i)$,
$\sum_i z_i =0$.  The authors of \GaiottoCD\ have studied the
analytic structure of the cross-ratios \formcross as a function of
$\zeta$ and have written a Riemann-Hilbert problem whose solutions
determine the metric in moduli space $g_{z_i \bar z_j}$.
 By computing the Kahler potential that leads
to this metric we can write an expression for the area as
\eqn\metricarea{
A \sim \sum_i (z_i \partial_{z_i} + \bar z_i \partial_{\bar z_i}) K ~,~~~~~~~~~
\partial_{z_i} \partial_{
\bar z_j} K = g_{z_i \bar z_j }
}
In principle this should determine the full solution of the
problem. The  metric is known explicitly for the case that
corresponds to four dimensional ${\cal N}=2$ theory with a single
hypermultiplet compactified on a circle, which have been
considered in \refs{\OoVa,\SeibergNS}. In this case the metric is
a multi-Taub-Nut metric. This corresponds to a case with only one
complex modulus, which leads to the octagon in our case \foot{We thank Davide Gaiotto for pointing out this relation.}.

\newsec{The octagon}

In the case of the octagon we have $n=4$ and the polynomial is $p
\sim z^2 - m $.
 As $\sqrt{p(z)}=z-{m \over 2z}+...$, we cover the $w-$plane twice,
  but, in addition we undergo a shift $w \rightarrow w+w_s$,
   with $w_s=-i \pi m$, as we go around twice. We can then
   say that the $w-$plane is missing a sliver of ``width" $w_s$.
    The fact that the information about $m$ survives at large $|w|$
     has the nice consequence that it allows us to compute the space-time
     cross-ratios exactly as a function of $m$
\eqn\singlc{ \eqalign{
\chi^+ \equiv & e^{ \wshifted + \bar \wshifted } =  e^{ \pi ( { m - \bar m \over i } ) } =
{ ( x_4^+ - x_1^+)  ( x_3^+ - x_2^+)
 \over ( x_4^+ - x_3^+ ) (x_2^+ - x_1^+)  }
\cr
\chi^- \equiv & e^{ \wshifted - \bar \wshifted \over i } =  e^{ - \pi ( { m + \bar m  } ) } =
 { ( x_4^- - x_1^-)  ( x_3^- - x_2^-)
 \over ( x_4^- - x_3^- ) (x_2^- - x_1^-)  }
}} We can now apply the known explicit formulas for the metric to
compute the area. The area \area\ is divergent and needs to be
regularized. We consider a physical regularization which
corresponds to placing a cut-off on the radial $AdS_3$ direction,
$r \geq \mu $.
 This cut-off renders the area finite since it does not allow arbitrarily large values of $|z|$ or $|w|$.

In order to extract the dependence on the regulator it is
convenient to write the area as the sum of two pieces
\eqn\regarsep{ \eqalign{ A = & 4 \int d^2z  ( e^{2\alpha} - \sqrt{
p \bar p} ) + 4 \int_{r(z,\bar z) \geq \mu } d^2z \sqrt{p \bar p}
= A_{Sinh} + 4 \int_{\Sigma} d^2 w \cr ~~~~&~{\rm with}~~~A_{Sinh}
= 4 \int d^2z ( e^{2 \alpha} - \sqrt{ p \bar p} ) = 4 \int d^2 w
(e^{2 \hat \alpha} -1) } } In order to regulate the second term in
\regarsep\ we need to know the asymptotic behavior of the radial
coordinate $r(z,\bar{z})$. This appears to require a full explicit
solution to the problem. However, we also know that the asymptotic
form of the solution also determines the positions of the cusps,
which in turn determine the kinematic invariants such as the
distance between cusps. Indeed, most of the dependence on   the
explicit solution $r(z,\bar{z})$ can be reexpressed in terms of
the kinematic invariants. We then find that
 the second piece in \regarsep\ can be written as the sum of several terms
\eqn\secpiece{4 \int_{\Sigma} d^2 w=A_{div}+A_{BDS}-{1 \over 2} \log(1+\chi^-)\log(1+{1 \over \chi^+}) -{ \pi \over 2}
  ( { m + \bar m} ) \log \stokesparameter_1^L -  { \pi \over 2 }   { m - \bar m \over i } \log \stokesparameter_1^R}
The first term is the well known divergent piece with the appropriate infra red behavior.
  $A_{BDS}$ is the function that appears at one loop in perturbation theory
\BernIZ, which solves the anomalous conformal ward identities
\DrummondAUA ,
 \eqn\bdsans{
 A_{BDS} =  - { 1 \over 4 } \sum_{i=1}^n \sum_{ j=1,   j \not = i,i-1 }^n  \log { x^+_j - x_i^+ \over x^+_{j+1} - x^+_i }
 \log { x^-_j - x^-_{i-1} \over x^-_j - x^-_i }
 }
 The last term in \secpiece\ depends explicitly on the Stokes
 parameters and is a consequence of the sliver missing in the
 $w-$plane. The explicit values of the Stokes parameters are given in
  \GaiottoCD
\eqn\gammaone{ \eqalign{
\log \stokesparameter_1 (\zeta)  =  { e^{ - i \phi} \zeta \over \pi } \int_{-\infty}^\infty dt { e^{ t }
\over e^{ 2 t } + \zeta^2  e^{- i 2 \phi } }  \log\left( 1 + e^{ -2 |m| \pi \cosh t }  \right)
\cr
 \log \stokesparameter_1^L = \log \stokesparameter_1(\zeta =1) ~,~~~~~~~~~~ \log \stokesparameter^R_1 = \log \stokesparameter_1(\zeta=i)
 ~,~~~~~~m = |m| e^{i \phi}
 }}
Finally,   $A_{Sinh}$ in \regarsep\ can be computed by considering the metric in moduli space
 for this problem \refs{\OoVa,\SeibergNS,\GaiottoCD}
\eqn\metricf{
 \partial_{m} \partial_{\bar m} K =    g_{m \bar m}  \sim
 \sum_{n=-\infty}^{\infty} {1 \over \sqrt{|m|^2   +(n+1/2)^2 } }  + {\rm const}
}
and using \metricarea .
Putting all this together we obtain the final answer for the null octagon
\eqn\finalan{ \eqalign{
 A = &  A_{div} + A_{BDS } + R
 \cr
 R = & -{ 1 \over 2} \log(1 + \chi^-) \log ( 1 + { 1 \over \chi^+ } )
+
  { 7 \pi \over 6 } +  \int_{-\infty} ^{\infty}
 dt  { |m| \sinh t  \over \tanh(2 t + 2 i \phi) }
 \log\left( 1 + e^{ - 2 \pi |m| \cosh t } \right)
 }}
 we have written the final answer in terms of the $BDS$ expression \bdsans\
  plus a remainder function, $R$,  which
 is a function of the cross ratios.
 For the special kinematical configuration
 we have considered in this note, this remainder
 functions depends on two cross ratios (out of a
  total of nine for the generic case in four dimensions). One way
  to think of this kinematic configuration is in terms of four
  left moving gluons $p_i^+$ and four right moving ones with
  $p_i^-$, $i=1,\cdots, 4$ in an $R^{1,1}$ subspace.
   Then $x^\pm_{i+1,i} = p^\pm_i$.
  This expression has the correct behavior for all the limiting
  cases and the correct analytic structure. It is analyzed in much greater detail in \AM .
Two loop perturbative
   expressions for the Wilson null polygons were given in \Brand .

 { \bf Acknowledgments }

We are very grateful to Davide Gaiotto, for a large number of important suggestions,
for his detailed explanations of \GaiottoCD\ and for giving us a draft of \GMNtwo .
We also thank N. Arkani-Hamed, V. Kazakov, P. Vieira  for discussions.

This work   was  supported in part by U.S.~Department of Energy
grant \#DE-FG02-90ER40542.

\listrefs

\bye